# Injection locking of GHz-frequency surface acoustic wave phononic crystal oscillator

Zichen Xi, Hsuan-Hao Lu, Jun Ji, Bernadeta R. Srijanto, Ivan I. Kravchenko, Yizheng Zhu, and Linbo Shao*

Z. Xi, J. Ji, Y. Zhu, L. Shao
Bradley Department of Electrical and Computer Engineering
Virginia Tech
Blacksburg, VA 24061, USA
Email: shaolb@vt.edu

Z. Xi, H. Lu
Quantum Information Science Section
Computational Sciences and Engineering Division
Oak Ridge National Laboratory
Oak Ridge, TN 37831, USA

B. R. Srijanto, I. I. Kravchenko
Center for Nanophase Materials Sciences
Oak Ridge National Laboratory
Oak Ridge, TN 37831 USA

Y. Zhu
Center for Photonics Technology
Virginia Tech
Blacksburg, VA 24061, USA

L. Shao
Center for Quantum Information Science and Engineering (VTQ)
Department of Physics
Virginia Tech
Blacksburg, VA 24061, USA

## Abstract

Low-noise gigahertz (GHz) frequencies sources are essential for applications in signal processing, sensing, and telecommunications. Surface acoustic wave (SAW) resonator-based oscillators offer compact form factors and low phase noise due to their short mechanical wavelengths and high quality (*Q*) factors. However, their small footprint makes them vulnerable to environmental variation, resulting in their poor long-term frequency stability. Injection locking is widely used to suppress frequency drift of lasers and oscillators by synchronizing to an ultra-stable reference. Here, we demonstrate injection locking of a 1-GHz SAW phononic crystal oscillator, achieving 40-dB phase noise reduction at low offset frequencies and unperturbed low noise at large offset frequencies. Compared to a free-running SAW oscillator, which typically exhibits frequency drifts of several hundred hertz over minutes, the injection-locked oscillator reduces the frequency deviation to below 0.35 Hz. We also investigate the locking range and oscillator dynamics in the injection pulling region. The demonstrated injection-locked SAW oscillator could find applications in high-performance portable telecommunications and sensing systems.

## 1. Introduction

Low-noise microwave-frequency oscillators are crucial components in telecommunication, instrumentation, navigation, and sensing systems. Photonic [1-8], optomechanical [9-12], integrated electronic [13-16] and acoustic-wave devices have been developed to achieve low-phase-noise microwave-frequency oscillators. Specifically, surface acoustic wave (SAW) devices based on piezoelectric materials such as aluminum nitride (AlN) [17-21], gallium nitride (GaN) [22-23], zinc oxide (ZnO) [24], and lithium niobate (LN) [25-29] enable efficient bidirectional conversion between electromagnetic waves and mechanical (acoustic) waves, making them promising platforms for developing compact, energy-efficient microwave oscillators with exceptionally low phase noise. Gain in SAW devices can be introduced by the acoustoelectric effect [30] and coupling to electronic components via transducers [31], and thus enabling simpler integration of the entire oscillation system in a single chip package. In addition, since high-quality ($Q$)-factor resonators are essential for low-noise oscillation systems, SAW resonators offer a distinct advantage by delivering significantly higher $Q$ factors than their electrical counterparts. For example, SAW resonators have achieved $Q$-factors of up to $10^5$ at gigahertz frequencies while occupying less than $300 \times 300$ μm$^2$ [28]; meanwhile, the most advanced non-superconducting electronic resonator exhibits $Q$ on the order of several hundred [16].

Capitalizing on high transmission efficiency and exceptional $Q$-factors, acoustic-wave oscillators - either employing acoustic wave delay lines or resonators - have shown competitive performance. Acoustic-wave oscillators operating at carrier frequencies in the megahertz to gigahertz range have been demonstrated on thin-film lithium niobate (LN) on insulator [24], suspended [17, 20, 22, 25-27] and bulk-substrate [29] structures, featuring a state-of-the-art phase noise of -132.5 dBc/Hz at a 10-kHz offset frequency of a 1 GHz oscillation frequency [29]. While acoustic-wave oscillators deliver superior short-term stability, the intrinsic temperature coefficient of frequency (TCF) of the materials renders them highly sensitive to environmental variations, resulting in large long-term frequency drift that limits their performance in practical applications. On the other hand, while some electronic oscillators may offer inferior short-term stability, their reduced sensitivity to temperature and humidity fluctuations guarantees greater reliability over prolonged periods. For example, a typical off-the-shelf voltage-controlled oscillator at gigahertz (for example, Mini-Circuits ROS-1121V+) features phase noise -110 dBc at 10-kHz offset frequency, while their phase noise at offset frequencies below 1 kHz could achieve -80 dBc/Hz, which are usually tens-of-dB better than the free-running SAW oscillators. Thus, using a signal from a long-term stable electronic oscillator to lock an acoustic-wave oscillator through the injection-locking mechanism could significantly improve the long-term stability of acoustic-wave oscillators, while maintaining the ultralow phase noise at high offset frequency. The injection locking could be useful in stabilizing oscillator-based sensors [32-33] against drifting over time.

In this work, we investigate injection locking, which is a widely exploited technique for frequency stabilization [7, 34-35], phase synchronization [36-37], and clock recovery [38] in communications and electronic systems, as a promising solution to enhance the long-term stability of our GHz SAW oscillators. We note that our work advances early explorations [39-40] of injection locking in 10s-MHz acoustic-wave oscillators by following aspects: (1) our device operates at much higher 1 GHz frequency and features lower phase noise, addressing needs of modern microwave applications; (2) we characterized the phase noise spectra with and without injection locking signal, revealing the phase noise performance; and (3) we provide a concise yet accurate equation of motion describing the oscillation loop, derive the locking range and noise suppression ratio, giving a clear mathematics and physics picture.

## 2. GHz-frequency SAW PnC Oscillator and Experimental Setup

We design and fabricate a SAW phononic crystal (PnC) resonator and subsequently use it to establish a self-oscillation system which features a simple and reconfigurable configuration. To achieve injection locking of our SAW oscillator, we inject a weak external microwave signal at a frequency within the locking range into the self-oscillation loop. Employing this injection locking mechanism, we experimentally demonstrate a 1-GHz injection-locked PnC SAW oscillator that achieves a phase noise of -80 dBc/Hz at the 30-Hz offset frequency, showing a significant improvement of 40 dB over its free-running counterpart. At higher offset frequencies that are beyond the locking range, the injection-locked oscillator follows the excellent short-term stability of its free-running counterpart, achieving a phase noise of -120 dBc/Hz at the 10-kHz offset frequency. We also observe a remarkable improvement in long-term frequency stability. Over a 200-second interval, the free-running oscillator experiences a frequency drift exceeding 600 Hz, whereas the injection-locked oscillator drifts by less than 0.35 Hz, showing an improvement of more than three orders of magnitude. Overall, our injection-locked SAW PnC oscillator features competent long-term and outstanding short-term frequency stability, making it an ideal integrated reference source for applications that demand high sensitivity and reliability, such as high-performance portable radar, satellite communications and sensing systems.

Figure 1(a) illustrates our SAW resonator configuration, which is formed by a series of etched grooves as PnC structures on a 128° Y-cut LN substrate. The acoustic wave is propagating in the X direction of LN crystal. A pair of interdigital transducers (IDTs) are used to convert electrical signals into acoustic waves and vice versa: the side IDT outside of the etched grooves region is positioned on the LN surface, the mid IDT inside the groove region is placed on the unetched surface between etched grooves. The design and fabrication details of such SAW PnC resonators are discussed in our previous works [28-29]. The constructed SAW resonator exhibits a high-transmission mode at 1025.68 MHz with a measured $S_{21}$ transmission of 2% (-17 dB) and a loaded $Q$ of ~2,300, as shown in Fig. 1(b), which is utilized to achieve the low-phase-noise self-oscillation.

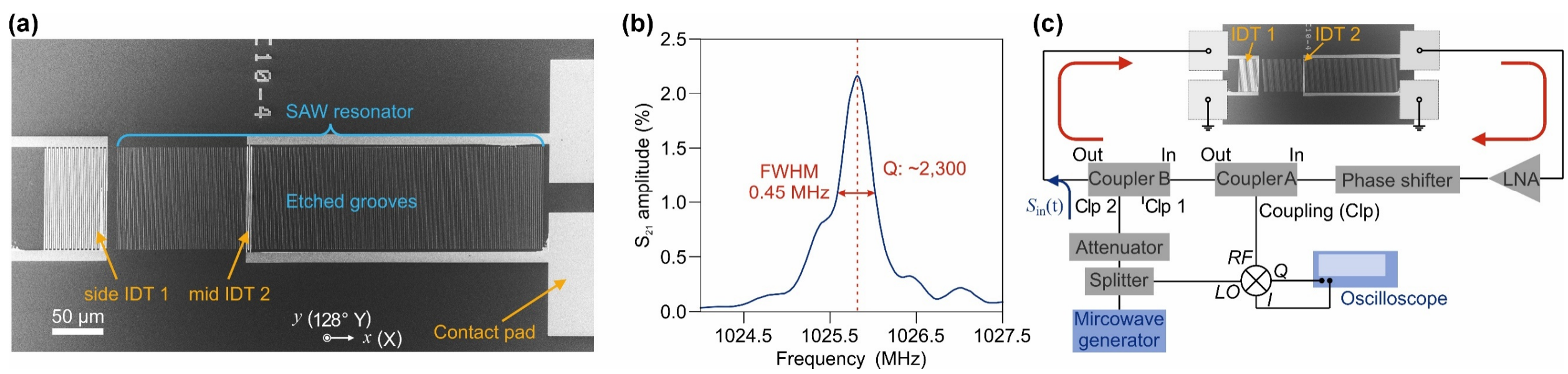

**FIG. 1.** (a) Configuration of the surface acoustic wave (SAW) resonator. The SAW resonator is defined by etched grooves. A pair of interdigital transducers (IDTs) is used to convert energy between electrical and mechanical fields. (b) Measured $S_{21}$ amplitude spectrum near the resonant mode frequency at 1,025.68 MHz, exhibiting a loaded quality factor ($Q_l$) of ~2,300 and a transmission of 2.16% (-17 dB) (c) Schematic of the injection-locked SAW oscillator measurement setup. The oscillator consists of the SAW resonator, a low-noise amplifier (LNA), a phase shifter and two couplers. An in-phase/quadrature ($I/Q$) demodulator and a high-resolution oscilloscope are used to capture the signal coupled out from coupler A for further data processing. A microwave signal generator, connected via a power splitter, provides the injected signal through coupling port 2 (Clp 2) of coupler B, and serves as the local oscillator (LO) for the $I/Q$ demodulator. A series of attenuators are used to fine-tune the injected power to the SAW oscillator.

The SAW oscillator consists of our SAW resonator, an LNA (Mini-circuits ZKL-33ULN-S+), a phase shifter (RF-LAMBDA RFPSHT0002W1) and 2 microwave couplers (Fig. 1(c)). The oscillator signal is measured from the coupling port of Coupler A (Mini-circuits ZFDC-10-5-S+). The coupling Port 1 of Coupler B (Mini-circuits, ZABDC20-182H-S+) is terminated with a 50-Ω resistor throughout the measurements. To record the oscillator signal for further analysis, we employ an in-phase/quadrature (I/Q) demodulator (Analog Device ADL5380) in conjunction with a high-resolution oscilloscope (Rohde & Schwarz RTO6). Additionally, we employ a microwave signal generator (Keysight N5183B with low phase noise option, phase noise -139 dBc/Hz at a 10 kHz offset of a 1 GHz signal) locked to a Rubidium frequency standard (Stanford Research Systems FS725, Allan variance $< 2 \times 10^{-12}$ at 100 second period) as a low-noise microwave signal reference. We split the electrical signal from the microwave signal generator into two paths: one path injects the signal into the oscillator loop via coupling Port 2 of Coupler B, while the other serves as the local oscillator (LO) for the I/Q demodulator. The output power of the signal generator remains fixed at 7 dBm throughout the measurements while a series of fixed attenuators are used to adjust the injected power into the SAW oscillator.

### 3. Injection Locking Dynamics

While previous works analyzed the injection locking problems based on Adler's equation [41-42] or rate equations [43-45], we analyze the dynamics of our SAW oscillator with injection signal using the following equation of motion based on the time-domain coupled mode theory [46-47], which provides a more general yet straightforward approach to address injection locking problems.

$$\frac{d}{dt}a(t) = -i\left(\omega_0 - i\frac{\gamma}{2} + ig_{NL}\big(a(t)\big)\right)a(t) + \sqrt{\kappa_{e1}}S_{in}(t), \tag{1}$$

where $a(t)$ is the amplitude of the mechanical resonant mode, in the unit of square root of phonon number. $\omega_0$ is the eigenmode angular frequency of the resonant mode. $\gamma = \frac{\omega_0}{Q_l}$ is the total loss of the mechanical mode, $Q_l$ is the loaded quality factor, and $\kappa_{e1}$ is the external coupling rate through interdigital transducer 1 (IDT 1). $S_{in}(t) = \tilde{S}_{in}e^{-j\omega_{in}t}$ is the injected signal with a magnitude of $\tilde{S}_{in}$ and an angular frequency of $\omega_{in}$. In our system, the nonlinearity is mainly contributed by the saturable electronic amplifier and all other components including our SAW resonator are working in their linear region; thus, the saturable gain can be written as

$$g_{NL}\big(a(t)\big) = \frac{\sqrt{\kappa_{e1}\kappa_{e2}}A_sL_c}{\sqrt{1 + \left|\frac{\sqrt{\kappa_{e2}}a(t)}{a_{sat}}\right|^2}} \approx \frac{\sqrt{\kappa_{e1}}A_sL_ca_{sat}}{|a(t)|}. \tag{2}$$

Here, $\kappa_{e2}$ is the external coupling rate through IDT 2, $A_s$ is the small-signal amplification factor of the low noise amplifier (LNA), $L_c$ is the loss of the passive components in the loop, and $a_{sat}$ is the saturation magnitude referenced to the input of LNA. The approximation holds when the LNA is in the deep saturation region, i.e. the signal at LNA input is much larger than the saturation magnitude, $\left|\sqrt{\kappa_{e2}}a(t)\right| \gg a_{sat}$.

When the LNA is deeply saturated and the oscillator is injection-locked, the oscillator frequency $\omega_{\text{osc}}$ will be synchronized to the injection frequency $\omega_{in}$, and thus we can rewrite $a(t) = \tilde{a} \cdot e^{-j\omega_{in}t}$, where $\tilde{a}$ is a constant complex constant. Taking into Eq. (1), we can obtain a quadratic equation about $|\tilde{a}|$,

$$\left(\frac{\gamma^2}{4} + (\omega_0 - \omega_{in})^2\right)|\tilde{a}|^2 - \gamma\sqrt{\kappa_{e1}}A_sL_ca_{sat}|\tilde{a}| + \left(\kappa_{e1}A_s^2L_c^2a_{sat}^2 - \kappa_{e1}\left|\tilde{S}_{in}\right|^2\right) = 0. \tag{3}$$

Assuming the injection signal is much weaker than the signal in oscillator (i.e. $\tilde{S}_{in} \ll A_s L_c a_{sat}$), we solve the discriminant ($\geq 0$) of Eq. (3) and obtain the locking condition $|\omega_0 - \omega_{in}| \leq \frac{\Delta_L}{2}$, where the locking range $\Delta_L$ is given by,

$$\Delta_L = \frac{\gamma|\tilde{S}_{in}|}{A_s L_c a_{sat}} = \frac{\omega_0}{Q_l L_c}\sqrt{\frac{P_{in}}{P_{LNA}}} \; . \tag{4}$$

Here $P_{\mathrm{LNA}}$ is the saturated output power of the LNA and $P_{in}$ is the power of the injected signal coupled into the oscillation loop. This locking range is consistent with previous works [41-45].

While Eq. (4), as the discriminant of Eq. (3), provides a necessary condition for the locking range, we can verify that it indeed defines the actual locking range by deriving Adler's equation from Eq. (1). Since the phase of $\tilde{S}_{in}$ can be absorbed into $a(t)$, without losing generality, we can write $S_{in}(t) = |\tilde{S}_{in}|(t)e^{-j\omega_{in}t}$. Further using $a(t) = \tilde{a}(t)e^{-j\omega_{in}t}$, Eq. (1) becomes

$$\frac{d\tilde{a}(t)}{dt} = j(\omega_{in} - \omega_0)\tilde{a}(t) + \left(\frac{\sqrt{\kappa_{e1}}A_s L_c a_{sat}}{|\tilde{a}(t)|} - \frac{\gamma}{2}\right)\tilde{a}(t) + \sqrt{\kappa_{e1}}|\tilde{S}_{in}|. \tag{5}$$

For steady state, $\tilde{a}(t) = \tilde{a}_0$, which results in Eq. (3). Writing in polar form, $\tilde{a}(t) = r(t)e^{j\theta(t)}$, where $\theta(t) = \theta_{in}(t) - \theta_{osc}(t)$ is the phase difference between the injected signal $\tilde{S}_{in}$ and output oscillation signal, the real and imaginary parts of Eq. (5) are

$$\frac{dr}{dt} = \left(\frac{\sqrt{\kappa_{e1}}A_s L_c a_{sat}}{r} - \frac{\gamma}{2}\right)r + \sqrt{\kappa_{e1}}|\tilde{S}_{in}|\cos\theta \, , \tag{6}$$

and

$$\frac{d\theta}{dt} = (\omega_{in} - \omega_0) - \frac{\sqrt{\kappa_{e1}}|\tilde{S}_{in}|}{r}\sin\theta \, . \tag{7}$$

For weak injection, $r(t)$ is near constant and $|\tilde{S}_{in}|$ is negligible. Therefore Eq. (6) leads to $r = 2\sqrt{\kappa_{e1}}A_s L_c a_{sat}/\gamma$. Substitute into Eq. (7), we have

$$\frac{d\theta}{dt} = (\omega_{in} - \omega_0) - \frac{\Delta_L}{2}\sin\theta \, , \tag{8}$$

which is the Adler's equation and produces the same locking range in Eq. (4). When locking, $d\theta/dt = 0$, and $\sin\theta = 2(\omega_{in} - \omega_0)/\Delta_L$.

To explore the noise behavior of injection-locked oscillators, we add noise terms into Eq. (8) by $\theta \mapsto \theta + \delta\theta$, $\omega_{in} \mapsto \omega_{in} + \delta\omega_{in}$, and $\omega_0 \mapsto \omega_0 + \delta\omega_0$, and get

$$\frac{d\delta\theta}{dt} = (\delta\omega_{in} - \delta\omega_0) - \frac{\Delta_L}{2}\sin(\theta + \delta\theta) + \frac{\Delta_L}{2}\sin\theta \, , \tag{9}$$

where $\delta\theta$ is the small phase deviation from the steady solution $\theta$. $\delta\omega_{in}$ and $\delta\omega_0$ are stochastic (noise) variables associated with injecting signal and the resonant mode, respectively. We define the corresponding phase noise terms as $\delta\theta_{in}$ and $\delta\theta_0$ such that $d\delta\theta_{in}/dt = \delta\omega_{in}$ and $d\delta\theta_0/dt = \delta\omega_0$. When $|\delta\theta| \ll 1$, $\sin(\theta + \delta\theta) \approx \sin\theta + \cos\theta \cdot \delta\theta$, we have

$$\frac{d\delta\theta}{dt} + \frac{\Delta_L}{2}\cos\theta \cdot \delta\theta = \frac{d\delta\theta_{in}}{dt} - \frac{d\delta\theta_0}{dt}. \tag{10}$$

We write Eq. (10) in Fourier domain as,

$$\left(j\omega + \frac{\Delta_L}{2}\cos\theta\right)\Theta(\omega) = j\omega\big(\Theta_{in}(\omega) - \Theta_0(\omega)\big), \tag{11}$$

where $\Theta(\omega)$, $\Theta_{in}(\omega)$, and $\Theta_0(\omega)$ are the Fourier transformation of $\delta\theta(t)$, $\delta\theta_{in}(t)$, and $\delta\theta_0(t)$.

Reorganizing Eq. (11), it gives

$$\Theta(\omega) = \frac{j\omega}{j\omega + \frac{\Delta_L}{2}\cos\theta}\left(\Theta_{in}(\omega) - \Theta_0(\omega)\right). \tag{12}$$

From $\delta\theta = \delta\theta_{in} - \delta\theta_{\mathrm{osc}}$, we have $\Theta(\omega) = \Theta_{in}(\omega) - \Theta_{\mathrm{osc}}(\omega)$, and Eq. (12) can be written as

$$\begin{aligned}\Theta_{\mathrm{osc}}(\omega) &= \Theta_{in}(\omega) - \frac{j\omega}{j\omega + \frac{\Delta_L}{2}\cos\theta}\left(\Theta_{in}(\omega) - \Theta_0(\omega)\right), \\ &= \frac{\frac{\Delta_L}{2}\cos\theta}{j\omega + \frac{\Delta_L}{2}\cos\theta}\Theta_{in}(\omega) + \frac{j\omega}{j\omega + \frac{\Delta_L}{2}\cos\theta}\Theta_0(\omega).\end{aligned} \tag{13}$$

Since phase noise of injecting signal $\delta\theta_{in}$ and resonant mode $\delta\theta_0$ are independent, we can rewrite Eq. (13) in power spectral densities (PSD),

$$S_{\theta_{osc}}(\omega) = \frac{\left(\frac{\Delta_L}{2}\cos\theta\right)^2}{\omega^2 + \left(\frac{\Delta_L}{2}\cos\theta\right)^2}S_{\theta_{in}}(\omega) + \frac{\omega^2}{\omega^2 + \left(\frac{\Delta_L}{2}\cos\theta\right)^2}S_{\theta_0}(\omega), \tag{14}$$

where $S_{\theta_{osc}}(\omega)$, $S_{\theta_{in}}(\omega)$, and $S_{\theta_0}(\omega)$ are the phase noise PSD of injection-locked oscillator, injecting signal, and the free-running oscillator. In other words, the phase noise PSD of the injection-locked oscillator is the weighted average between that of the injecting signal and the free-running oscillator.

In the limit of a good injecting signal ($S_{\theta_{in}}(\omega) \ll S_{\theta_0}(\omega)$), the noise suppression ratio is given by,

$$\frac{S_{\theta_{osc}}(\omega)}{S_{\theta_0}(\omega)} = \frac{\omega^2}{\omega^2 + \left(\frac{\Delta_L}{2}\cos\theta\right)^2}. \tag{15}$$

Eq. (15) indicates that a larger locking range $\Delta_L$ and smaller $\theta$ can lead to larger noise suppression ratio. This suppression ratio agrees with our experimental results, see following sections.

## 4. Experimental Characterization of Injection-Locked SAW PnC Oscillator

We experimentally characterize the SAW oscillator under both free-running and injection-locked conditions. Looking at a short time frame of 0.5 second, the oscillation frequency fluctuates with a peak-to-peak amplitude of approximately 80 Hz when free running and < 10 Hz when injection locked (Fig. 2(a)). The relative frequencies are extracted from the measured I/Q data by fitting the phase change with a period of 100 μs. Figure 2(b) shows the phase noise spectra of our oscillator, which are extracted by Fourier transformation of measured I/Q data. At the 30-Hz (1000-Hz) offset frequency, the injection-locked oscillator features a phase noise of -80 dBc/Hz (-103 dBc/Hz), which is 40-dB (13-dB) noise suppression. At offset frequencies above 6,000 Hz, the phase noise in both conditions is at a similar level, indicating that injection locking does not have a discernible impact on the SAW oscillator's phase noise floor at high offset frequencies (short time scale). Fig. 2(c) shows the noise suppression ratio, which is the difference in phase noises between the injection-locked oscillator and free-run oscillation. As the injecting signal features a much better phase noise than our free-running oscillator (Fig. 2(b)), the measured noise suppression agrees with the that of the good injecting signal limit (Eq. (15)).

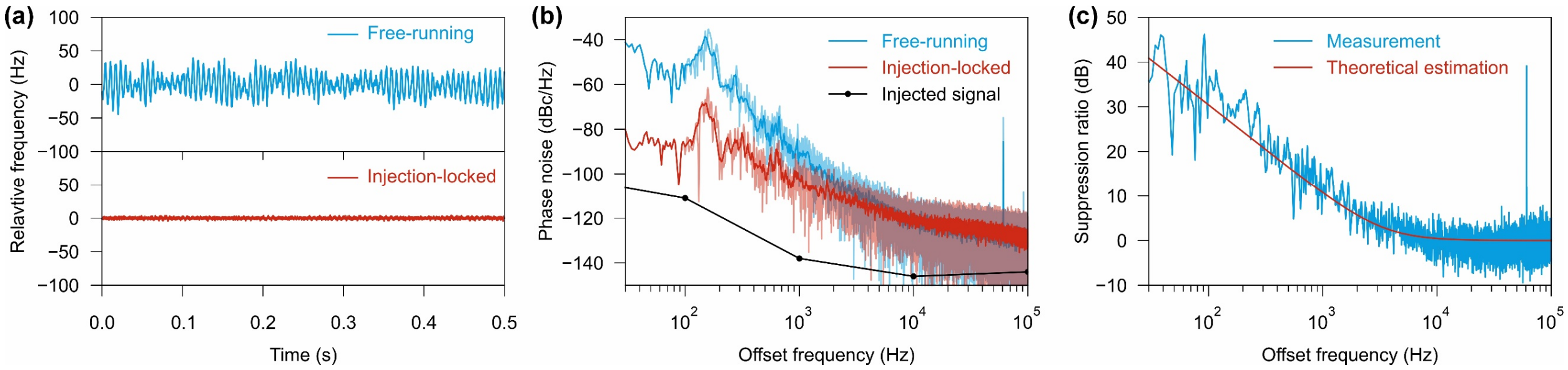

**FIG. 2.** (a) Relative frequency of the SAW oscillator under free-running and injection-locked conditions. The time resolution is 100 μs. The frequency is referenced to the external microwave signal generator backed by an atomic clock. (b) Measured phase noise diagrams of the free-running and injection-locked SAW oscillator. Curves in light (dark) colors represent raw data (moving averages). Both plots are extracted from the measured I/Q signals captured for 0.5 second. Black dots are phase noise of the injecting signal from the datasheet of the signal generator. (c) Phase noise suppression ratio by the injection locking. Experimentally measured (blue), and $\Delta_L$ = 6.6 kHz and $\theta = 0$ are used for theoretical estimation (red) based on Eq. (15).

We also test the long-term frequency stability of our SAW oscillator using a frequency counter (Fig. 3). The free-running oscillator drifts approximately 600 Hz over a 200-seconds measurement period. Meanwhile, the injection-locked oscillator maintains a stable carrier frequency (locked) with fluctuations less than 0.35 Hz (Inset in Fig. 3), demonstrating remarkable improvement in long-term stability of the SAW oscillator. We note that this 0.35-Hz frequency fluctuation is much smaller than that in the 0.5-seconds measurement with ~10-Hz fluctuations (Fig. 2(a)), it's because we use different time resolutions in these two set of measurements. The time resolution is 100 μs and 100 ms for measurements in Fig. 2(a) and Fig. 3, respectively.

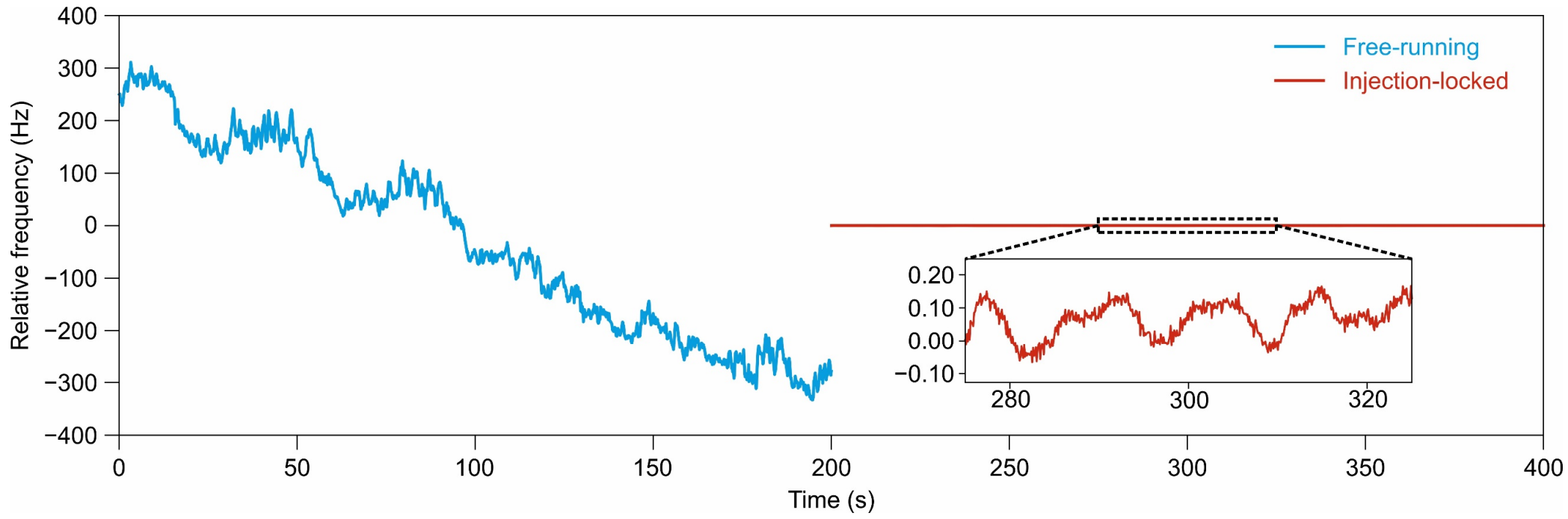

**FIG.3.** Long-term frequency-stability measurement of surface acoustic wave (SAW) oscillators under free-running and injection-locked conditions. A frequency counter is used to measure the oscillator frequency. The gate time for each measurement point is 100 ms. Inset: zoom-in view of the injection-locked oscillator.

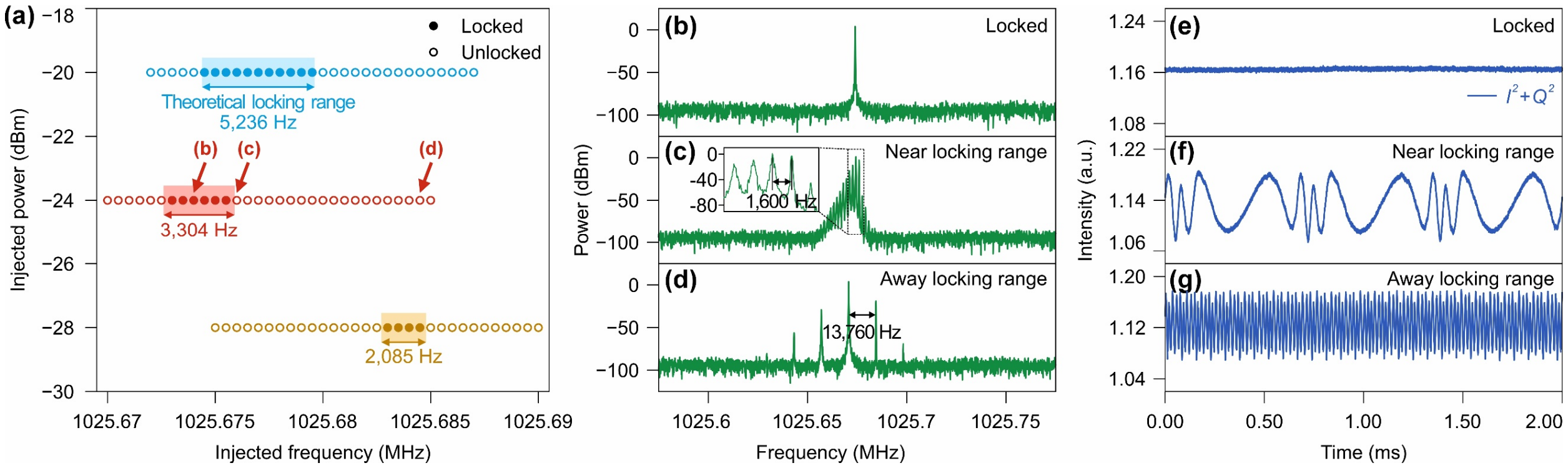

**FIG. 4.** Locking range characterization of the surface acoustic wave (SAW) oscillator. (a) Injection frequency sweep at three different injected power levels. Solid points (hollow circles) represent locked (unlocked) cases. The frequency step between two data points is 500 Hz, and the shaded blocks indicate the estimated locking range using Eq. (4). (b-d) The frequency spectra of the oscillator signal under injection-locked, near-injection-locked, and far-injection locked conditions, respectively. Inset: zoom-in view of the near-injection-locked spectrum. (e-g) The waveforms of average power levels corresponding to (b-d) extracted from the in-phase/quadrature data using $I^2 + Q^2$.

We further experimentally investigate the locking range under different injection power from -28 to -20 dBm (Fig. 4(a)). We note that the intrinsic resonant frequency $\omega_0$ is drifting due to environmental fluctuation, thus the measured locking ranges are not aligned in Fig. 4(a), and we use a sweep step of 500 Hz for the injection frequency to complete the sweep measurement in tens of minutes (limited by the data transfer from the oscilloscope). We measure locking ranges of 5000, 2500 and 1500 Hz at injected power of -20, -24 and -28 dBm, which agree with the estimated locking ranges of 5236, 3304 and 2085 Hz using Eq. (4). The device-related parameters of $f_0$ = 1025.68 MHz, $Q_l$ = 2,300, $L_c$ = 0.803, and $P_{LNA}$ = 112.2 mW are used. These differences between measured and predicted locking range are primarily due to the 500-Hz sweep step and the drifting of the intrinsic resonant frequency.

We show the frequency spectra of the oscillator signal under three injection conditions: injection-locked (Fig. 4(b)), near the locking range (Fig. 4(c)), and away from locking range (Fig. 4(d)), all with the same injected power of -24 dBm. The corresponding waveforms of the intensity of the oscillator output signals are shown in Figs. 4(e)-4(g). The scenarios shown in Figs. 4(c) and 4(d) are commonly referred to as injection pulling [7, 43-45]. When injection locking is achieved, the oscillator produces a single, narrow spectral line in the frequency domain, and maintains a quasi-constant average power in the time domain (Fig. 4(e)). On the other hand, injection pulling results in multiple spectral lines (Figs. 4(c) and 4(d)) and the signal intensity experience periodic waveforms (Figs. 4(f) and 4(g)). The frequency spacing of spectral lines are 1,600 (Inset of Fig. 4(c)) and 13,760 Hz (Fig. 4(d)). The periods in Fig. 4(f) and 4(g) also reflect the frequency difference between the intrinsic resonant frequency $\omega_0$ and the locking frequency $\omega_{in}$. We note that despite the spectra are reminiscent of a frequency comb, the intensity waveforms are different from that of a frequency comb [8].

## 5. Conclusion and Outlook

In summary, we have investigated injection locking of an SAW oscillator that demonstrates improved long-term frequency stability. Small-footprint acoustic-wave oscillators are notorious for frequency drift over time, but by leveraging injection locking, we improve long-term stability by more than three orders of magnitude without sacrificing short-term performance. Our SAW resonator is fabrication-friendly using standard processes and can be readily scaled to other operating frequencies. Moreover, our approach is compatible with conventional electronics and can be integrated within a small volume, which holds strong potential for applications in portable radar, telecommunication, and sensing systems.

## Conflict of Interest

The authors have no conflicts to disclose.

## Acknowledgement

We thank Prof. Honghu Liu for fruitful discussions on the mathematics of injection locking dynamics. Device fabrication and SEMs were conducted as part of user projects (CNMS2024-B-02643, CNMS2025-A-02944) at the Center for Nanophase Materials Sciences, which is a DOE Office of Science User Facility. This work was performed in part at Oak Ridge National Laboratory, operated by UT-Battelle for the U.S. Department of Energy under Contract No. DE-AC05-00OR22725, funding was in part provided by ORNL's Laboratory Directed Research and Development Program. This work was supported in part by the Defense Advanced Research Projects Agency (DARPA) OPTIM program under contract HR00112320031. This work was sponsored in part by the Air Force Office of Scientific Research (AFOSR) and was accomplished under Grant Number W911NF-23-1-0235. The views and conclusions contained in this document are those of the authors and do not necessarily reflect the position or the policy of the Government. No official endorsement should be inferred. Approved for public release; distribution is unlimited.

## Author Contributions

**Zichen Xi:** Investigation (lead), Methodology (supporting), Writing - Original Draft (lead). **Hsuan-Hao Lu:** Investigation (supporting), Writing - Review & Editing (lead), Funding acquisition (supporting). **Jun Ji:** Investigation (supporting), Methodology (supporting). **Bernadeta R. Srijanto:** Investigation (supporting). **Ivan I. Kravchenko:** Investigation (supporting). **Yizheng Zhu:** Investigation (supporting), Methodology (supporting). **Linbo Shao:** Conceptualization (lead), Methodology (lead), Investigation (supporting), Writing - Review & Editing (supporting), Supervision (lead), Funding acquisition (lead).

## DATA AVAILABILITY

The data that supports the findings of this study are available within the article. The raw data for the plots are openly available in figshare at http://doi.org/10.6084/m9.figshare.28862303. Other data that supports the findings of this study are available from the corresponding author upon reasonable request.